\documentclass[final,5p,times,twocolumn,authoryear]{elsarticle}

\usepackage[T1]{fontenc}

\usepackage{amssymb}
\usepackage{color}
\usepackage{geometry}
\usepackage{ulem}
\usepackage{soul}
\usepackage{caption}
\usepackage{amsmath,amsthm}
\usepackage{amsmath}

\usepackage{multirow}

\newcommand{\D}{\mathbb D}%
\newcommand{\E}{\mathbb E}%
\newcommand{\R}{\mathbb{R}}
\newcommand{\N}{\mathbb{N}}
\newcommand{\Pbb}{\mathbb P}%
\newcommand{\Mcal}{\mathcal{M}}%
\newcommand{\Pcal}{\mathcal P}%
\newcommand{\Ucal}{\mathcal U}%

\usepackage{chngcntr}

\begin{document}

\begin{frontmatter}


\author{François Lavallée\corref{co}\fnref{label1,label3}}
\ead{francois.lavallee@irstea.fr}
\author{Charline Smadi \fnref{label1,label3}}
\author{Isabelle Alvarez \fnref{label1,label3}}
\author{Björn Reineking \fnref{label2}}
\author{François-Marie Martin \fnref{label2}}
\author{Fanny Dommanget \fnref{label2}}
\author{Sophie Martin \fnref{label1,label3}}

\fntext[label1]{IRSTEA UR LISC, Laboratoire d'ing\'enierie pour les Syst\`emes Complexes, 9 avenue Blaise-Pascal CS 20085, 63178 Aubi\`ere, France}
\fntext[label3]{Complex Systems Institute of Paris Île-de-France, 113 rue Nationale, 75013, Paris, France}

\fntext[label2]{Univ. Grenoble Alpes, Irstea, LESSEM, 38000 Grenoble, France}

\cortext[co]{Corresponding author}

\title{A stochastic individual based model for the growth of a stand of Japanese knotweed including mowing as a management technique.}

\author{}

\address{}

\begin{abstract}

Invasive alien species are a growing threat for environment and health. They also have a major economic impact, as they can damage many infrastructures. The Japanese knotweed (\textit{Fallopia japonica}), present in North America, Northern and Central Europe as well as in Australia and New Zealand, is listed by the World Conservation Union as one of the world's worst invasive species. So far, most models have dealt with how the invasion spreads without management. This paper aims at providing a model able to study and predict the dynamics of a stand of Japanese knotweed taking into account mowing as a management technique. The model we propose is stochastic and individual-based, which allows us taking into account the behaviour of individuals depending on their size and location, as well as individual stochasticity. We set plant dynamics parameters thanks to a calibration with field data, and study the influence of the initial population size, the mean number of mowing events a year and the management project duration on mean area and mean number of crowns of stands. In particular, our results provide the sets of parameters for which it is possible to obtain the stand eradication, and the minimal duration of the management project necessary to achieve this latter.

\end{abstract}

\begin{keyword}
Invasive plant \sep  Fallopia spp. \sep  Reynoutria spp.\sep  Polygonum spp. \sep  individual based model \sep  management strategies \sep  dynamics \sep  model exploration
\end{keyword}

\end{frontmatter}

\section{Introduction}

Invasive alien species are a growing problem for environment and health. They may cause a loss of biodiversity \citep{murphy2014meta}, changes in ecosystem functioning \citep{strayer2012eight} or affect human well-being \citep{shackleton2019role}. They also have a major economic impact 
\citep{kettunen2008technical, pimentel2005update}. The necessity to act against invasive species relies on their global and local impacts and also on international policy engagements. For example, since 1992, the Convention on Biological Diversity (article 8\footnote{https://www.cbd.int/convention/articles/default.shtml?a=cbd-08}) compels the parties to "prevent the introduction of, control or eradicate those alien species which threaten ecosystems, habitats or species". Invasive species management raises several issues and strategies depend on the species, the step reached on the invasion process and the scale of action \citep{simberloff2013impacts}.

Together with field experiments, mathematical models can provide a better understanding of the critical determinants of the growth and spread of these species and thus help to identify efficient management strategies. Many studies focus on optimal management of invasive species \citep{baker2016placing, harris2009invasive, travis2011improving}. Control strategies aim at bringing the density of the invasive species below a threshold. For seed dispersal species, a frequent question in the literature of invasive species through mathematical modelling is to assess the benefit of a spatial prioritization: is it more profitable to remove the individuals at the heart of the infection, or those on the periphery \citep{harris2009invasive} ? The answer depends essentially on the spatial spread of the plant (and therefore on the species considered). Another issue in the literature of invasive species management is the temporal distribution of the effort: is it better to act significantly at the beginning of the management project and then to control the invasion with a lower effort (as in \cite{meier2014space}), or to increase the effort over time (as in \cite{baker2016placing})? A common assumption when dealing with control strategies is that the invasive species has already been present for a long time \citep{baker2016placing}. We know however that early detection and rapid response to the invasion may have a greater efficiency \citep{pyvsek2010invasive}.


Among the worst invasive species threatening biodiversity, Asian knotweeds raise particular management issues. This complex of three species (the Japanese knotweed, \textit{Fallopia japonica} [Houtt.] Ronse Decraene, the giant knotweed, \textit{Fallopia sachalinensis} [Schmidt Petrop.] Ronse Decraene and the hybrid between the two previous the Bohemian knotweed (\textit{Fallopia $\times$ bohemica} Chrtek \& Chrtkova) have invaded Europe and North America. Native to Eastern Asia, knotweeds have been introduced for ornamental purpose at the end of the 19th century \citep{bailey2004distribution,barney2006biology,beerling1994fallopia}. They are also present in Australia, New Zealand and Chile \citep{alberternst2006invasive,saldana2009fallopia}.

 
Asian knotweeds quickly invade the environment in which they grow \citep{gowton2016influence} and have large impacts \citep{lavoie2017impact}. They displace other plant species through light competition and allelopathy  \citep{dommanget2014differential, siemens2007evaluation}, affect native fauna diversity  \citep{abgrall2018invasion, gerber2008exotic, maerz2005green, serniak2017effects} and modify ecosystem functioning \citep{dassonville2011niche,tharayil2013phenolic}. In addition, the control costs are very high and were estimated at 250 million dollars a year in Great Britain \citep{colleran2014situ} and more than 2 billion euros a year in Europe \citep{kettunen2009technical}. 


Asian knotweeds grow in a wide variety of soils: sandy, swampy, rocky. They mainly invade human modified habitats such as roadsides, waste dumps, but also river banks. They are perennial geophytes: their rhizomes allow them to spend the winter season buried in the ground \citep{de2001viability}. Their rhizomes also play a major role in their propagation, thanks to their strong regeneration capacities \citep{bailey2009asexual, brock1992regeneration}. Once arrived in a new area, the rhizome expands centrifugally and a new stand can sustainably establish in a few weeks \citep{gowton2016influence,smith2007simulation}.


Once established and due to their extensive rhizome network, Asian knotweeds are extremely hard to remove. Rhizomes represent two third of their biomass \citep{barney2006biology} and can expand several meters away from the visible invasion front \citep{barney2006biology}. The resources they store can be efficiently remobilized after mowing events \citep{rouifed2011contrasting}. Some authors estimate that six cuttings are needed to significantly reduce belowground biomass \citep{gerber2010evaluating}. Understanding the underground development of Asian knotweeds is crucial to gain insight in their local propagation and performance. Moreover it could help to better design efficient management strategies.


As underground organs are almost inaccessible to observers, direct observations are scarce and models could help to approach their dynamics and better understand how management actions can affect their development. To our knowledge, there are very few models in the literature that describe the growth of a Japanese knotweed stand, and among them, rare are those that include a management technique. 

In \cite{smith2007simulation}, the authors build a 3D correlated random walk model of the development of the subterranean rhizome network for a single stand of Japanese knotweed. Their model is based on knowledge of the morphology and physiology of the plant. They study the model through simulations and they observe a quadratic expansion of the area invaded.

\cite{dauer2013elucidating} propose an "Integral Projection Model", inspired from matrix population models, for the plant dynamics at the level of a stand. The variable of interest is a continuous variable which stands for the height or the total biomass of the plant, and the authors use a simplified plant life cycle to model transition between states, like the transition from new shoots to crown (a crown is the location of a terminal bud from which stems emerge). They study the parameters that have the largest effect on the growth rate of the population.

\cite{gourley2016mathematical} develop a mathematical model for biocontrol of \textit{Fallopia japonica} using one of its co-evolved natural enemies, the Japanese sap-sucking psyllid \textit{Aphalara itadori}. It is a deterministic model that describes the evolution of the number of insects (larvae and adults), the total weight of the knotweed stems and the rhizome biomass. A key parameter of their model is the duration that a larva takes to consume and digest the rhizome biomass of the plant.

A commonly used management technique for Asian knotweed stands is mowing. Managers can vary its intensity and frequency, which motivates a study of the effects of these two parameters on the dynamics of the stand. This paper aims at understanding the influence of mowing on the growth of a Asian knotweed stand. More precisely, we study the influence of the initial population size, the mean number of mowing events a year and the management project duration on mean area and mean number of crowns of the stands. To our knowledge, existing models are not well designed to study such questions. Here we present a stochastic individual based model for the growth of a stand of Japanese knotweed including mowing as a management technique. The stochastic formalism enables us to study the early stage of invasion, without assuming the species has been present for a long time. The description of phenomena at the level of individuals enables us to take into account the variability between crowns, for example due to different ages. The study will focus on the influence of management parameters on model outputs which are area and number of crowns of the stand.

The paper is organized as follows. Section \ref{Section_Materials_and_methods} is devoted to the description of the ecological mechanisms taken into account in the model (apical dominance, intraspecific competition, etc.), the presentation of the mathematical model, as well as the methods we used for our study. Results are given in Section \ref{Section_Results}. In particular, we performed a calibration of the plant dynamics with field data from stands observed in the French Alps. We also studied, using numerical simulations with OpenMOLE software \citep{Reuillon2013}, the influence of management parameters on the population growth. Finally, we summarize our results and discuss their implications and shortcomings in Section \ref{Section_Discussion}.

\section{Materials and methods}  \label{Section_Materials_and_methods}

In this section we provide a description of the dynamical model for the growth of a stand of Japanese knotweed including mowing as a management technique. We also describe the methods used to its study  through numerical simulations.

In the sequel, the term individual will refer to crowns, we recall that a crown is the location of a terminal bud from which stems emerge. Individuals are characterized by their position $x$ in the plane and their underground biomass $a$ (i.e. the biomass rhizome connected to the crown).

The following notations will be needed to describe our model:
\begin{itemize}
 \item $ \chi = \R^2 \times \R_+$, is the state space of positions and biomasses. 
In the model, a crown is represented by a Dirac mass $\delta_{(x,a)}$, with $(x,a) \in \chi$, where $x$ stands for the position of the crown and $a$ is the biomass associated with the crown. 

\item  The set of crowns present at time $t$ is described by the measure $Z_t \in \Mcal( \chi ) $, where  $\Mcal( \chi ) $ is the set of finite point measures on $ \chi$ whose masses of points are 0 or 1.  
\begin{equation*}
\Mcal( \chi ) = \left\{  \sum_{i=1}^n  \delta_{(x_i,a_i)} ~ , ~ n \geq 0 , ~ (x_1,a_1), \ldots, (x_n,a_n) \in      \chi  \right\}. 
\end{equation*}   
\item $\Mcal_F ( \chi )$  [resp. $\Pcal(\chi)$] is the set of finite measures (resp. probability measures) on $\chi$, such that $\Mcal( \chi ) \subset \Mcal_F ( \chi )$.
\end{itemize}

Using $\Mcal_F ( \chi )$  allows us not setting \textit{a priori} the number of individuals in the model, since it contains all the possible population sizes, $n\geq 0$.

\subsection{Description of the phenomena included in the model}
\label{Section_Description_Phenomena}

\noindent \textbf{Birth:} an individual with trait $(x,a) \in \chi$ (i.e. its position is $x$ and it has an underground biomass $a$) will give birth at the rate $b(x,Z)$, where $Z \in \Mcal( \chi )$ describes the state of the system (i.e. the positions and biomasses of all individuals). 
It is assumed here that this birth rate does not depend on $a$.
A crown can give birth to several crowns and the rates at which an individual gives birth takes into account the proximity with its neighbouring individuals.
Based on \cite{smith2007simulation}, we consider that a crown will give birth to at most  two daughter crowns. We introduce the real parameter $distanceParent$, so that the birth rate of a crown depends on the number of crowns that are at a distance smaller than $distanceParent$ from it. So, if a crown has already given birth to its two daughters (in fact if there are already three individuals which are at a distance smaller than $distanceParent$ from it since we count its parent), it will have a zero birth rate (it does not give birth anymore). We will see in the next paragraph that a daughter crown can in principle be at a distance greater than $distanceParent$ from its parent. This phenomenon may be balanced by the fact that crowns from another parent can be at a distance less than $distanceParent$.
This modelling allows us to account for the effects of apical dominance: if a crown dies, the apical dominance it exerts on the neighbouring lateral buds ceases, and these last ones may develop to form aerial shoots, and thus form a new crown.

\cite{bashtanova2009physiological}, \cite{adachi1996central} and \cite{dauer2013elucidating} mention this phenomenon of apical dominance in a general way, but they do not specify a typical distance. That is why we will use a calibration method to set its value (in fact we use this method for all parameters, cf.  Section \ref{SubSectionCalibration}).

The rate at which an individual  with position $x$ gives birth can thus be expressed as follows:
\begin{equation}
\label{eq_birth}
b(x,Z) = \bar{b}_{\{ \sum_{y \in V(Z) } 1_{ \{  |x-y|\leq distanceParent \} } ~ \leq 3  \}  },
\end{equation}
where $$V(Z) := \{ x \in \R^2, Z( \{x\} \times  \R_+ ) >0  \}$$ is the set of the positions of the crowns present in the population $Z$.

\bigbreak

\noindent \textbf{Dispersal of the created individual: }\label{paragraph_DispersalOfTheCreatedIndividual} an individual with trait $(x,a)$ which gives birth generates an individual at position  $x'$. 
Here we choose a Gamma law density denoted by $D$ for the birth distance law. The parameters ($shape$, $scale$) of the law will be subject to calibration. We assume a uniform distribution on $x'-x$ direction angle to the X-axis.

Moreover, we will model the phenomenon of intra-specific competition by considering that an individual is really born if it does not fall too close to an already existing crown (otherwise we consider that it is not born). So we introduce the set $C$ depending on the population state $Z$ and position of the potential parent $x$:

\begin{equation}
\label{ensemble_dispersion}
 C_{x,Z} = \{  z \in \R^2 ,  ~\forall y \in V(Z)\setminus \{x \}, ~~ |y-z|>distanceCompetition \}.  
\end{equation}

The individual created must therefore be at a distance larger than $distanceCompetition$ from its neighbours not to fall in the zone of intraspecific competition. This principle of excluded zones (for the birth of an individual) is also used in \cite{smith2007simulation}: the zones surrounding the crowns are subject to competition for light and the appearance of new crowns is not allowed. 

\begin{figure}[h]
    \centering
     \includegraphics[ width=0.80\linewidth]{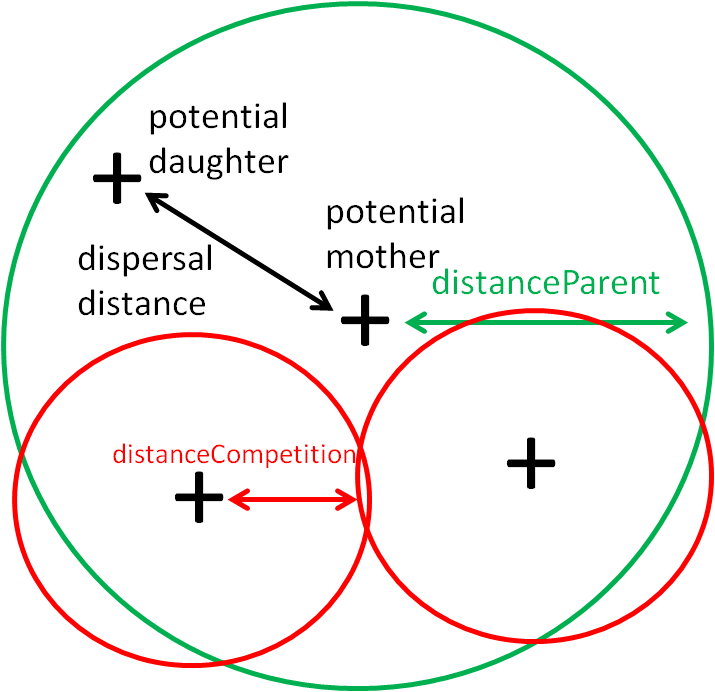}  
     \caption{Diagram representing a birth event. A cross stands for a crown position. We also indicate the different distances used in the model.}     
   \label{Figure_DiagramBirth}
\end{figure}

The diagram in Figure \ref{Figure_DiagramBirth} shows distances playing a role during a birth event.

\bigbreak

\noindent \textbf{Evolution of the biomass: } in \cite{seiger1997mechanical}, the authors present the effects of mowing on rhizome growth. They find that rhizome biomass increases significantly throughout the growing season, unlike above-ground biomass, which no longer grows significantly at the end of the summer. If the aerial shoots are not cut, the growth of the underground biomass $a$ is assumed to evolve according to a Von Bertalanffy's law \citep{paine2012fit}, presented below (Equation \eqref{VonBert}). The effects on rhizome development of mowing aerial shoots are not well known. We assume that mowing results in a decrease of underground biomass. This assumption stems from the fact that rhizome resources are used for aerial shoot regeneration (see \cite{gerber2010evaluating}). In \cite{rouifed2011contrasting}, the authors also note that mowing impacts the amount of underground biomass at the end of the season, and that mowing induces a decreasing rhizome density with depth (whereas without mowing it is constant). However, we do not take this phenomenon into account, 
since the model is planar.

Mowing events occur at a rate $1 / \tau$ (there is thus a mean of $\tau$ mowing events a year), and a proportion $proportionMowing$ (constant) of individuals is mown. 

After a mowing event, it is assumed that the underground biomass $a$ of an individual is immediately impacted and becomes $a.F(a)$, where $F$ takes values in $[0,1]$ and describes  the mowing effect as a function of the individual biomass. In order to take into account the fragility of young crowns, it is assumed that the function $F$ describing the impact of mowing on biomass is higher for low biomasses (we therefore take $F$ increasing, which implies that for two biomasses $a_1 < a_2$, we will have $ a_1 * F(a_1) < a_2 * F(a_2)$). 
We suppose that $F$ has the form:
\begin{equation} \label{fauche_lambda}
\forall a \in \R_+, ~F(a) = 1- exp(-mowingParameter  * a),
\end{equation} 
where $mowingParameter$  is a parameter that acts on the decay rate of the exponential function.

\bigbreak

We now have to specify how we choose the crowns to be mown. In this paper, we consider two possibilities. The first one is to choose the mowed crowns uniformly at random (random management technique). The proportion can thus represent different qualities of mowing if the aim is to mow the whole stand, with respect to different tools used (by hand, brush cutter). We consider this technique when we write the mathematical formalism of the model in Section  \ref{SectionMathFormalism}. The second one is mowing one side of the stand: it consists in determining an abscissa at the right of which every crown is mowed, and at the left of which no crown is mowed (side management technique). It reflects for example the case of a stand located on two plots, owned by different persons, one who manages the stand whereas the other does not. This situation occurs frequently along roadsides. We use this management technique in the model for the calibration due to the characteristics of our data set. 

\bigbreak 


For the biomass growth of a crown  when there is no mowing, we use Von Bertalanffy's Equation \eqref{VonBert}. First described in \cite{von1934untersuchungen}, this equation has then been very often used especially in forestry \citep{zeide1993analysis}. Here, this equation describes the evolution of the biomass as a function of time. It is based on simple physiological arguments: the growth rate of the organism decreases with biomass.

\begin{equation}
\label{VonBert}
\frac{da(t)}{dt}= L  ( K - a(t) ) =:  v( a(t)),
\end{equation}
where  $L$ is the growth rate at low biomass and $K$ is the ma\-ximal asymptotic biomass.

We can solve exactly Equation \eqref{VonBert}. If we suppose that the biomass at time $t_0$ is equal to $a_0$, then for all $t \geq t_0$, we get:
\begin{equation}
\label{solVonBert}
a(t) = a_0 e^{-L(t-t_0)} + K (1- e^{-L(t-t_0)} ). 
\end{equation}

Moreover, we assume that all crowns are born with the same biomass $a_0 \in \R^+$.


\bigbreak 

\noindent \textbf{Mortality:} we assume that the mortality rate is independent of the individual position. An individual alive at time $t$ and with biomass $a(t)$ dies at a rate $m(a(t))$. We assume that the mortality $m$ is a decreasing function of the biomass: an individual with a low biomass, either because it has just been created or because it has been mown, has a higher mortality rate. So if $T_0$ is the date at which an individual born at time $0$ dies, and whose biomass up to time $t$ is given by the function $a$ (we suppose it is not mown), we get:

\begin{equation}\label{proba_date_mort} 
\Pbb (T_0 \geq t) = e^{ - \int_0^t m(a(s)) ~ds}. 
\end{equation}

In \cite{smith2007simulation}, the authors set the probability that a segment of rhizome dies over a four month period (a time step in their model) to 0.0083.
When there is no mowing, mortality events for crowns rarely occur in nature, that is why the value proposed in \cite{smith2007simulation} is low. Having a good estimate for this value requires sufficiently large numbers of observations, thus calibration is very useful to estimate a value for such a parameter. We suppose that $m$, the function that describes the mortality rate of a crown according to its biomass, has the form:

\begin{equation} \label{eqtaux_mortalite}
m(a)= deathParameterScaling ~ e^{-\text{deathParameterDecrease} *a}.
\end{equation}  
Equation \eqref{eqtaux_mortalite} involves two parameters:   $deathParameterDecrease$, which influences the decay rate of the function and \textit{deathParameterScaling} which enables to choose the mortality rate for individuals with low biomass.


We summarize model parameters in \ref{Section_Summary_of_model_parameters}.
They are of two kinds: management parameters and plant dynamics parameters.

\subsection{Mathematical formalism associated with the model}
\label{SectionMathFormalism}

The class of stochastic individual-based models we are extending in this work were introduced by \cite{bolker1997using}, and by \cite{dieckmann2000geometry}. A rigorous probabilistic description and study was then conducted by \cite{fournier2004microscopic}. Since then, these models have been widely studied and extended (for instance in \cite{champagnat2006microscopic, champagnat2006unifying, costa2016stochastic, coron2018stochastic}).

The model proposed here and its mathematical study are drawn from the work of \cite{tran2006modeles, tran2008large} and  \cite{fournier2004microscopic}. In particular, notations and techniques derive from these papers.

We recall that a crown is represented by a Dirac mass $ \delta_{(x, a)} $, with $ (x, a) \in \chi $, where $x$  indicates the position of the crown and $a$ its biomass. The set of crowns  present in the population at time $ t \geq 0 $ is described by the measure $ Z_t \in \Mcal( \chi ) $.

The stochastic differential equation \eqref{equation_sto} describing the plant population dynamics is governed
by $ M_1 $, $ M_2 $ and $ M_3 $ three independent Poisson random measures, defined as follows:

\begin{itemize}
\item$M_1(ds,di,d\theta,dz)$ is a Poisson random measure on $ \R_+ \times \N^* \times \R_+ \times  \R^2 $ with intensity $ds \otimes n(di)  \otimes d\theta \otimes  D(dz) $, where $n(di)$ stands for the counting measure on $\N^*$ and $D$ is the density of the law for the dispersal of a child. The measure $M_1$ describes the birth events.

\item $M_2(ds,dy)$ is a Poisson random measure on  $ \R_+ \times [0,1]^{\N^*} $ with intensity $1 /\tau ~ ds \otimes \Ucal^{\N^*} ( [0,1])  $, where  $\Ucal ( [0,1]) $ is the uniform law on $[0,1]$. We denote  $y= (y_1,y_2,\ldots)$ for $y \in [0,1]^{\N^*}$. The measure $M_2$ describes the mowing events.

\item $ M_3(ds,di,d\theta) $ is a Poisson random measure on $ \R_+ \times \N^* \times \R_+   $  with intensity $ds \otimes n(di)  \otimes d\theta   $, where  $n(di)$ stands for the counting measure on $\N^*$. The measure $M_3$ describes the death events.
\end{itemize}

\bigbreak 

In Equation \eqref{equation_sto} below, $$Z_t = \sum_{i=1}^{N_t} \delta_{( X_i(Z_t) , A_i(Z_t) )}$$ is thus the measure that describes the population at time $t \geq 0$, $A_b$ is the flow of the differential equation describing the evolution of the rhizome biomass of a crown (Equation \eqref{solVonBert}).
$X_i$ (resp. $A_i$) denotes the position (resp. biomass) of the $i$-th individual in the population (in lexicographical order).
Let functions $b : (x,Z) \in \R^2 \times \Mcal_F(\chi) \mapsto b(x,Z)$  and $m : a \in \R_+ \mapsto m(a)$ be respectively individual birth and death rates.
The application $$ C : Z \in  \Mcal_F(\chi) \mapsto C_{X_i(Z),Z} \in \Pcal (\R^2 )$$ gives the admissible region for the births of new individuals which is related to intraspecific competition.
The function $F : [0,K] \to [0,1]$ models the effect of mowing crowns and  $\tau$ is the average number of mowing events a year:

\begin{small}
\begin{equation}
\label{equation_sto} 
\begin{array}{lll}
Z_t  =  \sum_{i=1}^{N_0} & \delta_{( X_i(Z_0) , A_b(t,0,A_i(Z_0) )}  \\ 
  + \int_0^t \int_{\N^*} \int_{\R_+} \int_{\R^2} &   1_{ \{  i \leq  N_{s-}  \} } \delta _{( X_i(Z_s)+z , A_b(t,s, a_0 ) ) }  1_{ \{  \theta \leq b(X_i(Z_s), Z_s ) \} }  \\
 &  1_{ \{ X_i(Z_s) + z ~ \in ~ C_{X_i(Z_s),Z_s} \} } M_1(ds,di,d\theta,dz)   \\ 
 +  \int_0^t \int_{[0,1]^{N^*}} \int_0^1 ~ &   \sum_{i=1}^{N_{s^-}}  1_{ \{ y_i \leq proportionMowing \} } ~( \delta_{( X_i(Z_s) , A_b(t,s,A_i(Z_{s^-}).F(A_i(Z_{s^-}) ) ))}    \\  
 &  -  \delta_{( X_i(Z_s) , A_b(t,s,A_i(Z_s)) )} )   M_2(ds,dy)  \\  
 - \int_0^t \int_{\N^*} \int_{\R_+}  &  1_{ \{  i \leq  N_{s-}   \} }   1_{ \{  \theta \leq m( A_i(Z_s) ) \} }   \delta _{( X_i(Z_s) , A_b(t,s,A_i(Z_s) ) }  M_3(ds,di,d\theta)  
\end{array}
\end{equation}
\end{small}

\bigbreak

The first term in Equation \eqref{equation_sto} refers to the evolution of the initial population: those individuals keep their position cons\-tant but their biomass evolves with the flow $A_b$. As mentioned above, the second term refers to birth events. A birth event consists in choosing a potential parent and verifying whether it satis\-fies the conditions to give birth: it is the role of the indicator functions. If it occurs, we add a Dirac mass corresponding to a new individual in the population.  The middle integral term refers to the mowing event for which an individual artificially  dies and is replaced by an other individual with the same position and a reduced biomass.
The last term refers to death events, for which we delete an individual in the population subtracting a Dirac mass.

\bigbreak 
Under boundary conditions over the birth and death rates (let us denote by $\bar{b}$ the upper bound of $b$), we have the following result (obtained in a similar way as in \citep{tran2006modeles}, Propositions 2.2.5 and 2.2.6):
if $Z_0 \in \Mcal( \chi )$, the stochastic differential equation admits a unique pathwise strong solution $(Z_t)_{t \in \R_+} \in \D( \R_+ , \Mcal( \chi ) )$ such that for all  $T>0$, the number of individuals at time $t \leq T$ $N_t :=  \langle Z_t,1 \rangle = \int_{\R^2 \times \R} Z_t(dx,da) $ satisfies :
\[  \E[ \sup_{t \in [0, T]} N_t ] < \E [ N_0] e^{\bar{b}T} < \infty .   \]

This gives in particular an upper bound to the growth of the population when there is no management.

\subsection{Simulation of the model} 

The algorithm used to simulate a solution of the stochastic differential equation \eqref{equation_sto} is presented in \ref{Subsection_Description_algo}.
To illustrate the evolution of the stand under our model, we use the software Scala (version 2.11.12).
We use OpenMOLE software (\cite{Reuillon2013}, version 8.0) to perform the model exploration.
Finally, we use R software (version 3.4.4) for the statistical analysis of model outputs.
Simulations were performed on the European Grid Infrastructure (http://www.egi.eu/).

\begin{figure}[h]
    \centering
     \includegraphics[ width=0.90\linewidth]{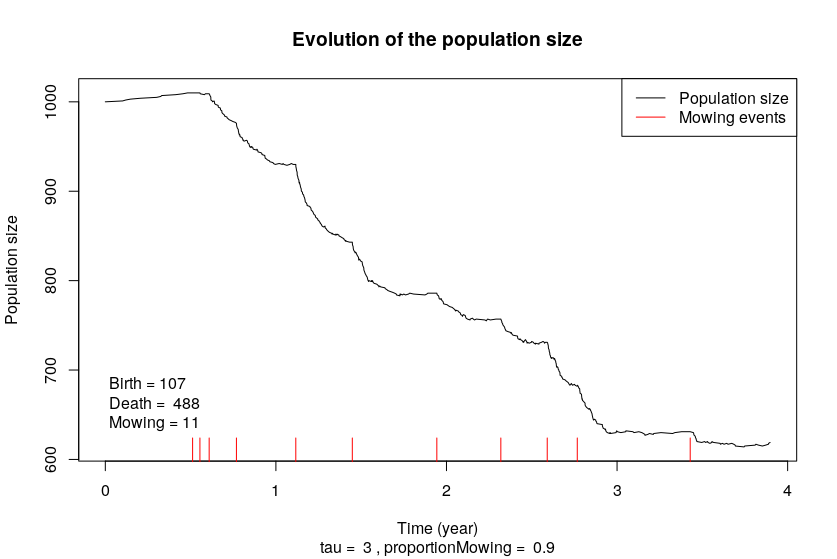}  
     \caption{Simulation of one trajectory of the model (Equation \eqref{equation_sto}) with $\tau = 3$, initial population size = 1000, $T =4$ and $proportionMowing = 0.9$ with plant dynamics parameters from Table \ref{tableau_result_Calibration}. Black line shows the population size, red lines indicate dates of mowing events.}
   \label{Figure_OneSimulation}
\end{figure}

Figure \ref{Figure_OneSimulation} illustrates the evolution of the population size (number of crowns) of one trajectory of the model, for given parameters of the plant dynamics and management parameters. The initial population size was set to 1000, the mean number of mowing events a year $\tau = 3$, the management project duration $T=4$ years and the proportion of mown crowns $proportionMowing = 0.9$. We thus have a mean number of mowing events equal to $3 \times 4= 12$ (there were 11 in the simulation). The final population size is equal to 619, so the management strategy leads to a reduction of roughly one third in population size.  

\subsection{Data for the Calibration}  %
\label{SubSectionCalibration}


Our goal is to find the influence of management parameters on the stand dynamics. We must therefore set parameters of the plant dynamics. For some of the parameters, we could not find values in the literature. We thus proceed to a calibration which consists in finding parameter values of the plant dynamics with which the model best reproduces field data. 

The field data are those used by \cite{martin2018Biological}.
The authors studied the invasion potential of the Japanese knotweed along an elevational gradient (i.e. in mountains), by identifying the determinants of its spatial dynamics. The experiment consists in collecting data on stands of Japanese knotweed at different altitudes in the French Alps. The measurements were carried out in 2008 and 2015, on the stands themselves (outline, stem density,...) as well as on biotic and abiotic variables. Stands were mown or not, and for each stand, we have access to some information about the management technique used by the land owner: the frequency of mowing and an estimation of the proportion of the mown stand, which corresponds to the side management technique. There is a high variability in the observed stands, both in size (from less than $ 2 m^2$ to $350 m^2$) and land conditions in which they grow such as soil quality, proximity of river, road, forest, abandoned land.

In model outputs, we compute the final and initial population sizes and areas (the area of the stand is the area of the convex hull formed around the simulated stands). From \cite{martin2018Biological}, we use data about stand areas and crown densities so we can deduce the population size, for stands in 2008 and 2015.

As mentioned in \ref{Subsection_Description_algo}, the simulation operation for a stand takes place in two stages: first, the creation of the initial population given a population size to reach (we choose the size of a stand in 2008), then its evolution, according to the information related to management techniques contained in data from \cite{martin2018Biological}.

\subsection{The method used for the calibration}

When we consider a set of parameters for the plant, and perform a simulation for each of the 19 stands, we obtain $19 \times 4$ results (real numbers): the areas and sizes of the initial and final populations. We can thus compare these values with the $19 \times 4$ corresponding observations of \cite{martin2018Biological}. The goal is to find a set of parameters for the plant, common to all stands, that best matches the model outputs (area and size) to field observations. Notice that there is in particular the comparison between the initial size of the observations of 2008 and the one that has been simulated. This comparison is mainly used to check if the set of parameters of the plant being tested makes it possible to obtain an initial population. Indeed, some sets of parameters can lead to a failure in the creation of populations (e.g. if the distance of competition is too large whereas the dispersal distance is too small).

We thus need a distance to compare the simulations and the observations. We have chosen here, for each stand and each type (area or size), the distance: $ dist(simu,data) = abs (simu - data) / data $. We have chosen to use a relative error distance (renormalization by the data) because areas and sizes have not the same order of magnitude, and there is also a wide difference within size values and area values themselves. 
The total distance to minimize is the sum of the distances over the 19 stands and over the 4 observations (size and area, in 2008 and 2015). Notice that if the set of parameters does not allow for the creation of an initial population, we obtain a population of size zero and a null area at the initial time (2008), and the simulated values for 2015 are also null. The distance between the observations and a trivial (null) population is equal to $76 = (19 \times 4)$.

In order to minimize this distance, the OpenMOLE software  proposes a method based on genetic algorithms for model cali\-bration (NSGA2). The result obtained is presented in Section \ref{Subsection_resultCalibration}. The calibration algorithm is an iterative algorithm, which provides at each step a set of solutions. As steps go by, the distance $dist(simu,data)$ between data and simulation results for the selected solutions decreases.

\subsection{Numerical analysis}  \label{Section_Numerical_analysis}
Simulations studied in the following are performed with the set of parameters for the plant dynamics obtained by calibration (Section \ref{Subsection_resultCalibration}, Table \ref{tableau_result_Calibration}).

Let us now explain how we studied the influence of the management parameters $\tau$ and $T$ and of the initial population size, where we recall that 
\begin{itemize}
\item $\tau$ is the mean number of mowing events a year
\item $T$ is the duration of the project
\end{itemize}

We focus on the influence of these three parameters and we do not study the influence of the $proportionMowing$ parameter. We set its value to $0.9$ and use the random management technique. 
Indeed, we consider that the manager aims at mowing the whole stand, but we do not use a value of $proportionMowing$ equal to one in order to consider an imperfect mowing due to the tool used (as mentioned in Section \ref{Section_Description_Phenomena}).


We perform samplings of management parameters in OpenMOLE, with a replication of size n=50 for each set of management parameters (these samplings are detailed in Section \ref{Subsection_Influence}) and we calculate the mean quantities over these $n$ values. Indeed, with our management point of view, we are interested in the mean behaviour of a stand. We first let one parameter vary. Based on an initial visual inspection of simulation results, we fit three relationships via least squares: a linear regression performed with $R$ function \textit{lm}, a truncated quadratic relationship (Equation \eqref{eqRegressionGlobaleArea}), and an exponential regression (Equation \eqref{eqRegressionGlobaleSize}) performed with $R$ function \textit{nls}. We assess model performance by the coefficient of determination ($R^2$) and the root mean squared errors (\textit{RMSE}).
Finally, in Section \ref{Section_Formulas}, we use the same statistical tools (\textit{lm} and \textit{nls}) to derive general regression formulas for the mean output quantities depending on management parameters, and two constants that the algorithm aims to find.

\section{Results}
\label{Section_Results}

\subsection{Calibration}
\label{Subsection_resultCalibration}

The values of calibrated parameters are presented in Table  \ref{tableau_result_Calibration}. 

The set of solutions provided by the NSGA2 algorithm stabilized after 165000 steps. For a set of parameters, \textit{evol.sample} refers to the number of replications that were carried out by the algorithm.  Since our model is stochastic, we need to choose a solution with a sufficiently large value for \textit{evol.sample}. Among the set of solutions provided by the algorithm, we chose the solution that was replicated at least 50 times, and that minimizes the distance $dist(simu,data)$.

\bigbreak
\begin{minipage}{1\linewidth}
\begin{tabular}{lll}
	\hline
        Variable			&  Value after calibration    & unit  \\
   	\hline
		K						&	12.72				  &	  g  \\	
		L						&	0.26  				  & $year^{-1}$   \\
		distanceCompetition		&	0.15 				  &	  m  \\	
		distanceParent			&	0.20 				  &	  m  \\
		shape					&	4.34 				  &	     \\	
		scale					&	2.36 				  &	     \\
		deathParameterDecrease	&	2.32 				  &	$g^{-1}$   \\ 	
		deathParameterScaling	&	1.12 				  & $year^{-1}$   \\
		mowingParameter			&	0.11 				  &	$g^{-1}$   \\	
		bbar					&	0.18 				  &	$year^{-1}$   \\
		a0						&	1.73 				  &	  g  \\	
		score 		&	26.06				  &      \\
		evolution.samples		&	79	 				  &	     \\	
     \hline                                   
\end{tabular}
 \captionof{table}{ Result of the calibration obtained with OpenMole software \citep{Reuillon2013}  }
 \label{tableau_result_Calibration}
\end{minipage}
\bigbreak

In Table \ref{tableau_result_Calibration}, $score$ is the median over the 79 replications of the sum of the  distances $dist(simu,data)$ over the 19 stands and the 4 characteristics (initial or final and area or size).
A score of 26.06 means that in half of the cases and on average, the relative distance for one characteristic between the simulated stand and the corresponding data is lower than 0.3.
The reason for this difference is that  data were obtained from field work that was not carried out in order to calibrate the model, and thus contain a bias due to the altitude or soil type.

Even though we could not find values for the plant dynamics parameters in the literature, experts can provide boundaries for some of them. Thus, we can assess the ecological quality of the result given by the algorithm. First, the parameters $distanceCompetition$ and $distanceParent$ are close to what is expected according to our field experience. Then the distribution for the dispersal of individuals is close to the one suggested by specialists (Figure \ref{Figure_GammaLaw}).
In Figure \ref{Figure_Mortality}, we plot the mortality rate of a crown according to its biomass. We note that a crown that is not mown keeps a very low mortality rate, in agreement with field observations. Indeed, to compare with the value in \cite{smith2007simulation}, we calculate with Equation \eqref{proba_date_mort} and the parameter values from calibration, the probability that a crown dies before 4 months. This quantity is equal to 0.0027, which has the same order of magnitude as the value found in \cite{smith2007simulation} for the probability of a rhizome segment dying in a four months period (0.0083).

Finally, the ratio between the value of $K$ (maximal biomass, that is likely to be found for the oldest crown, i.e. in the center of the stand when there is no mowing), and $a_0$, the biomass of a crown at  birth (rather in periphery) equals 7.4 (the ratio is expected to be around 10 in \cite{adachi1996central}).

\subsection{Influence of management parameters and initial population size}
\label{Subsection_Influence}

In this section, the aim is to find 
statistical relationships between the explanatory variables (management parameters or the initial population size) and model outputs (mean area and mean size of a stand).

We consider the two following samplings: 
\begin{itemize}
\item In $sampling1$, we make $\tau$ vary in $[0,15]$ in steps of 0.5, $T$ in $[0,16]$ in steps of 1 and $initialPopSize$ equals either 500 or 1500. We run 50 simulations of our stochastic model for each of these sets of values. $sampling1$ has a high sampling rate on $\tau$ and $T$, with high values for the initial population size. It is used in Sections \ref{Section_T} and \ref{Section_Tau} to study more precisely these two management parameters. 

\item In $sampling2$, we make $\tau$ vary in $[0,14]$ in steps of 2, $T$ in $[0,16]$ in steps of 2 and $initialPopSize$ in $[50,1200]$ in steps of 50. We run 50 simulations of our stochastic model for each of these sets of values.
$sampling2$ has a high sampling rate 
on the initial po\-pulation size, it is used in Section \ref{Section_TailleIni} to study more precisely its influence.
\end{itemize}

\subsubsection{Influence of management duration $T$} \label{Section_T}

In this section, we use the first sampling ($sampling1$) to study the influence of the management duration $T$ on the final mean areas and sizes. 
Given $\tau$ and a value of the initial population size, we perform 

\begin{itemize}
\item a truncated quadratic regression  for the mean final area.

\item a non linear regression on strictly positive values for the mean final population size, using the function $f(T) = InitialPopSize * exp(-T / rate) $, with $rate$ being a constant on which the algorithm \textit{nls} maximizes $R^2$.
This constant rate is different for each set of parameters, since it depends on the values of $\tau$ and $initialPopSize$. In Section \ref{Section_Formulas}, we study this dependency. 
\end{itemize}

\begin{figure}[h]
    \centering
     \includegraphics[ width=0.90\linewidth]{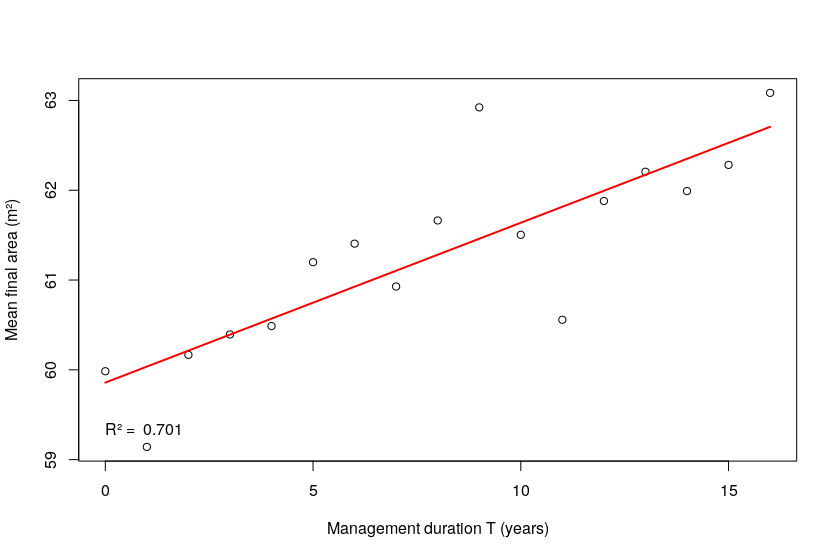}  
     \caption{Linear regression of mean final area as a function of management duration $T$, with $\tau = 0.5 ~ year^{-1}$ and initial population size = 1500.}
   \label{Figure_size_vs_T_lowTau}
\end{figure}

It turns out that for values of $\tau \leq 2.5$, the mean area remains close to its initial value at time $T=0$ (a maximum relative difference of $3 m^2$), and the variation is rather linear but the corresponding $R^2$ values are below 0.9. Figure \ref{Figure_size_vs_T_lowTau} gives an example of the linear regression on the mean area according to $T$, for given values of $\tau \leq 2.5$ and initial population size.

\bigbreak

\begin{figure}[h]
    \centering
     \includegraphics[ width=0.90\linewidth]{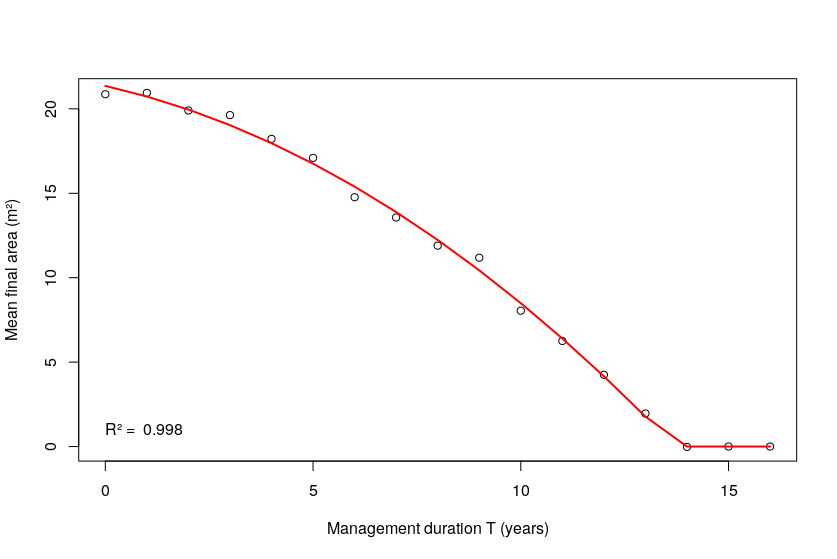}  
     \caption{Quadratic regression of mean final area as a function of management duration $T$, with $\tau = 8 ~ year^{-1}$ and initial population size = 500. For the last three points in the bottom right hand corner, at least half of the 50 simulations lead to extinction.
     }
   \label{Figure_area_vs_T}
\end{figure}

$R^2$  and \textit{RMSE} values enable us to conclude that the  mean final area depends quadratically on the management duration $T$ when $\tau> 2.5$. Indeed, 47 regressions out of the 50 in the sampling (variation of initial Population size and $\tau > 2.5$) lead to an $R^2$ value larger than 0.95. The maximal value of \textit{RMSE} over these 50 regressions is  1.31 $m^2$, which is low compared to the mean initial area which has values 20 $m^2$ or 60 $m^2$. Figure \ref{Figure_area_vs_T} gives an example of the quadratic regression on the mean area for given values of $\tau$ and initial population size.

\begin{figure}[h]
    \centering
     \includegraphics[ width=0.90\linewidth]{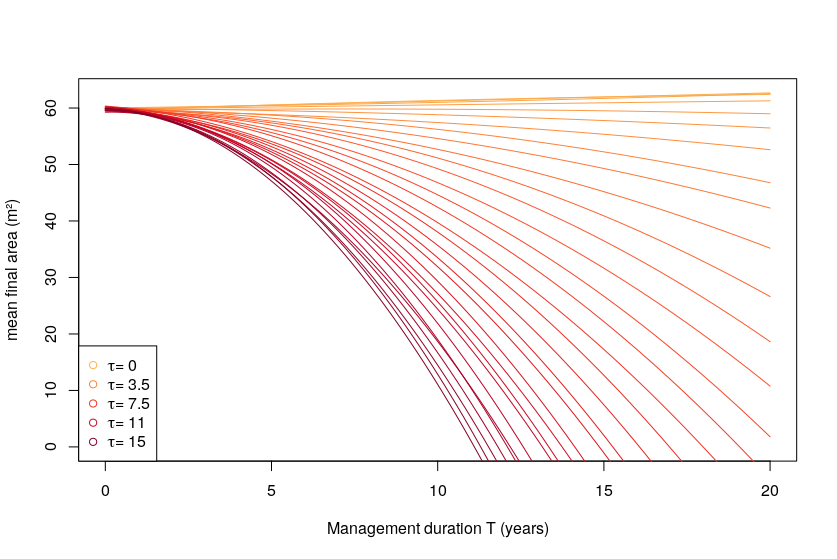} 
     \caption{Quadratic regression curves of the mean final area as a function of $T$, for different values of $\tau$ (low in clear colours, up to 15 mowings per year in dark colours), and we set $InitialPopSize = 1500$ and $proportionMowing = 0.9$.  }
   \label{Figure_QuadraticRegresionCurve}
\end{figure}

\bigbreak
These results give information about the influence of the duration of the management project on stand growth. A first fact, which is very important for management is that it is not sufficient to mow to decrease the population size and area: if the number of mowings per year is too low (less than 2.5 in our case), the population size and area increase during the management project.
In Figure \ref{Figure_QuadraticRegresionCurve}, we plot the quadratic regression curves for the average final area  with respect to the duration of the management project ($T$, on the abscissa), obtained for different $\tau$. 
A second important fact for management is that we cannot expect an eradication of a knotweed stand of initial area $60$ $m^2$ in less than 11 years (for 15 mowing events a year). The figure also tells us about the stand surface reduction in terms of final mean area when mowing once more time per year. For example, mowing 6 times a year instead of 5, during 10 years, reduces the final surface of the stand by $4$ $m^2$ on average (looking at the section $T=10$ on Figure \ref{Figure_QuadraticRegresionCurve}).

\bigbreak

As for the area, the size varies linearly for values of $\tau \leq 2.5$ ($R^2$ around 0.9). For larger values of $\tau$, we present here the result of the non linear regression. 

$R^2$  and \textit{RMSE} values enable us to conclude that the  mean final size depends exponentially on the management duration project $T$, when $\tau> 2.5$. Indeed, all the 50 regressions in the sampling (variation of initial Population size and $\tau > 2.5$) lead to an $R^2$ value larger than 0.95. The maximal value of \textit{RMSE} over these 50 regressions is 39 crowns, which is low compared to the initial population size which has values 500 crowns or 1500 crowns.

\begin{figure}[h]
    \centering
     \includegraphics[ width=0.90\linewidth]{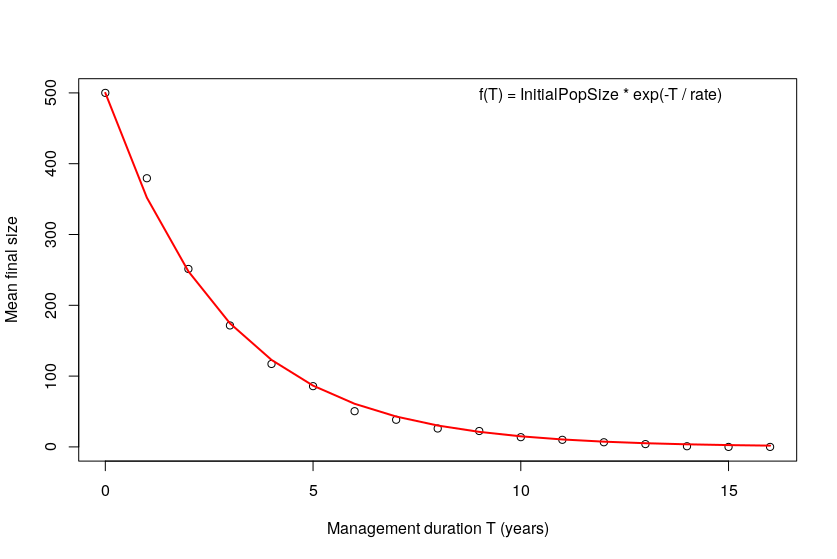}  
     \caption{Non linear regression of mean final size as a function of management duration project $T$ with $\tau = 8 ~ years^{-1}$ and initial population size = 500. For the last three points in the bottom right hand corner, at least half of the 50 simulations lead to extinction.
     }
   \label{Figure_size_vs_T}
\end{figure}

Figure \ref{Figure_size_vs_T} gives an example of the quadratic regression on the mean area for given values of $\tau$ and initial population size.

\subsubsection{Influence of the mean number of mowing events a year $\tau$}
\label{Section_Tau}

In this section, we use the first sampling ($sampling1$) to study the influence of the mean number of mowing events a year on the final mean areas and sizes. Given a value of initial population size and $T$, we perform 

\begin{itemize}
\item a linear regression on strictly positive values of outputs for the mean area.

\item a non linear regression on strictly positive values for the mean final population size, using the function $f(\tau) = InitialPopSize * exp(-\tau / rate) $, with $rate$ being a constant on which the algorithm \textit{nls} minimizes the sum of squared errors.  This constant rate differs for each set of parameters since it depends on the values of $T$ and $initialPopSize$. As mentioned for the similar constant in Section \ref{Section_T}, we study this dependency in Section \ref{Section_Formulas}.
\end{itemize}

The linear regression presented below holds for  $\tau > 2.5$ (as in Section \ref{Section_T}) and $T \geq 2$ (to have a decreasing population).

$R^2$  and \textit{RMSE} values enable us to conclude that the  mean final area depends linearly on the mean number of mowing events $\tau$. Indeed, 24 regressions over the 30 in the sampling (variation of initial Population size and $T \geq 2$) lead to an $R^2$ value larger than 0.95. The maximal value of \textit{RMSE} over these 50 regressions is $2.09$ $m^2$, which is low compared to the mean initial area which has values 20 $m^2$ or 60 $m^2$.

We also conclude that the  mean final size depends exponentially on the mean number of mowing events $\tau$. Indeed, all the 30 regressions in the sampling (variation of initial Population size and $T \geq 2$) lead to an $R^2$ value larger than 0.95. The maximal value of \textit{RMSE} over these 50 regressions is 64 crowns, which is low compared to the initial population size which has values 500 crowns or 1500 crowns.

\subsubsection{Influence of the initial population size}
\label{Section_TailleIni}

In this section, we use the second sampling ($sampling2$) to study the influence of the initial population size on the final mean areas and sizes. Given a value for $\tau$ and $T$, we perform a linear regression on strictly positive values of outputs. Due to the wide range of values for the initial population size in the $sampling2$, too many extinctions may occur for a given set of management parameters. We thus perform the regression only if there are at least 5 strictly positive output values. 

$R^2$  and \textit{RMSE} values enable us to conclude that both the  mean final area and the mean final size depend linearly on initial population size. Indeed, in both cases, 60 regressions over the 63 regressions  among the 72 management sets in the sampling lead to an $R^2$ value larger than 0.95. The maximal value of \textit{RMSE} over these 63 regressions is $1.1$ $m^2$ (resp. 18 crowns) for the mean final area (resp. the mean final size) case, which is low compared to the mean initial area (resp. initial population size) which ranges from 2 $m^2$ to 48 $m^2$ (resp. from 50 crowns to 1200 crowns).
 
Notice that the influence of the initial population size on the initial area is also linear. 
Indeed, the $sampling2$ contains the case $T=0$, and for this specific value of $T$ the final area is the initial area.

\subsubsection{Formulas for the mean final sizes and areas, as functions of $\tau$, $T$ and the initial population size}  \label{Section_Formulas}      

We summarize results of the regressions we performed in Table \ref{tableau_SummarizeInfluence}.


\bigbreak

\fontsize{8pt}{9pt}\selectfont
\begin{tabular}{lllll}
	\hline
     Parameter                  &    mean output    &  Variation                & $R^2$ > 0.95  &   RMSE \\
   	\hline
     $T$ ~~ for $\tau \geq 2.5$ &	final area		&  	quadratic   $\searrow$  &  47/50   &  1.31   \\
	 $T$ ~~ for $\tau \geq 2.5$ &	final size		&  	exponential $\searrow$  &  50/50   &  39   \\ 				 $\tau$ ~~~ for $\tau \geq 2.5$ &	final area		&  	linear 		$\searrow$  &  24/30   &  2.09   \\
 $\tau$ ~~~ for $\tau \geq 2.5$ &	final size		&  	exponential $\searrow$  &  30/30   &  64   \\
	 $InitialPopSize$   	    &	final area		&  	linear 		$\nearrow$	&  60/63   &  1.1   \\
	 $InitialPopSize$   	    &	final size		&  	linear 	    $\nearrow$  &  60/63   &  18   \\

	     \hline                                   
\end{tabular}
\captionof{table}{Summary of the regression results of Sections \ref{Section_T}-\ref{Section_TailleIni}.}
\label{tableau_SummarizeInfluence}
\normalsize

\bigbreak    

In Sections \ref{Section_T} to \ref{Section_TailleIni}, we have studied the influence of one parameter, while the two others were set constant. The two previous samplings introduced at the very beginning of this Section \ref{Subsection_Influence} were designed to control the variation of management parameters and initial population size, in order to investigate their influence on the model outputs. Based on results in Sections \ref{Section_T} to \ref{Section_TailleIni}, we are now able to propose a formula for the mean areas and sizes as a function of the two management parameters (the mean number of mowing events a year ($\tau$) and the management project duration ($T$)) and the initial population size. We use a Sobol sampling (that maximizes discrepancy of the sequence, i.e. the space is evenly covered) of 5000 points with $ \tau \in [0 ; 15.0]$, $T \in  [0 ; 20]$,  and $initialPopSize \in [100 ; 1500]$.

For the same reason as before, we consider the case of $\tau \geq 2.5$. Equations \eqref{eqRegressionGlobaleSize} and \eqref{eqRegressionGlobaleArea} highlight relationships between final outputs, management parameters and initial population size.
\begin{equation} 
\label{eqRegressionGlobaleSize}
Mean ~ Final ~  Size ~ = InitialPopSize \times exp(- T  .( \tau - a)/b ),  
\end{equation} 
with $a,b \in \R$  constants, and
\begin{equation}
\label{eqRegressionGlobaleArea}
Mean ~ Final ~  Area ~ = \max( (c \times \tau +d) \times T^2 + 0.04 \times InitialPopSize, 0) 
\end{equation} 
with $c,d \in \R$  constants.

\bigbreak
We now discuss the results of the non linear regression with respect to the two management parameters ($T$ and $\tau >2$) and initial population size (the sampling contains 4332 values for the triplet ($T$, $\tau$, $InitialPopSize$)). $R^2$ and \textit{RMSE} between the predicted values and the data for the mean size are respectively equal to 0.99 and 26.12 crowns. 95 \% confidence intervals for the constants $a$ and $b$ are given by \textit{R} and their values are: $a \in [ 0. 90 ; 0.94 ]$ and $ b \in [20.46 ; 20.77]$. $R^2$ and \textit{RMSE} between the predicted values and the data for the mean area are respectively equal to 0.99 and 2.23 $m^2$. Moreover, the corresponding 95 \% confidence intervals for the constants $c$ and $d$ are $c \in [ -0.0342 ; -0.0336]$ and $ d \in [0.0960 ; 0.0998]$, respectively. 
Taking $T=0$ in the right hand side of Equations \eqref{eqRegressionGlobaleSize} and \eqref{eqRegressionGlobaleArea}, gives $InitialPopSize$ and $InitialPopSize * 0.04$, respectively. The last quantity thus corresponds to the mean initial area. There is indeed a linear dependency between the mean initial area and the initial population size.

\begin{figure}[h]
    \centering
     \includegraphics[ width=0.90\linewidth]{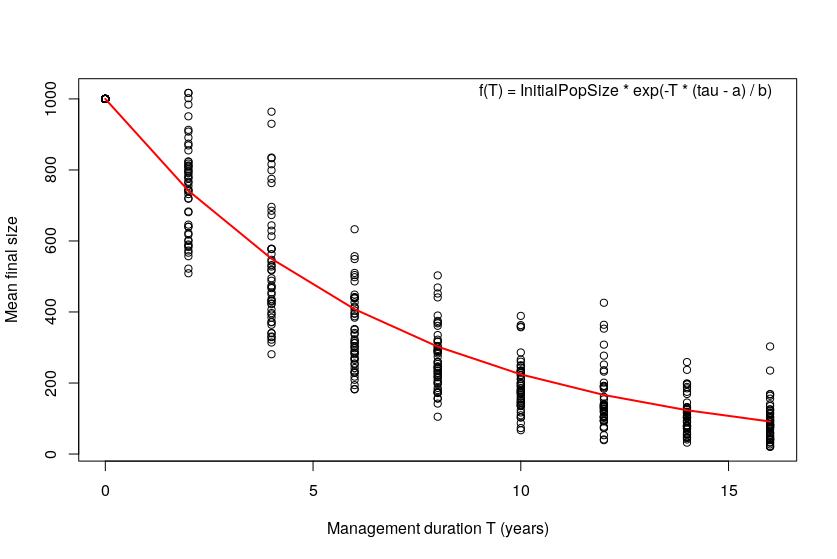} 
     \caption{Exponential regression for the mean final size as a function of the management duration project ($T$). The red line is the prediction function of $T$ defined by Equation \eqref{eqRegressionGlobaleSize}. Black circles represent stand sizes resulting of 50 replications with $\tau = 4$, initial population size = 1000, letting $T$ vary.}
   \label{Figure_Global_size_vs_T}
\end{figure}

An important remark on the generality of Equations \eqref{eqRegressionGlobaleSize} and \eqref{eqRegressionGlobaleArea} is the following: results obtained for the mean output quantities are still relevant for direct outputs, without considering mean quantities.  Figure \ref{Figure_Global_size_vs_T} illustrates this point, plotting the formula \eqref{eqRegressionGlobaleSize}, for given values of $\tau$ and initial population size and by letting $T$ vary. We emphasize that the red line on Figure \ref{Figure_Global_size_vs_T}  has been found with a regression on a far larger set of points than the subset selected to plot this example.

\begin{figure}[h]
    \centering
     \includegraphics[ width=0.90\linewidth]{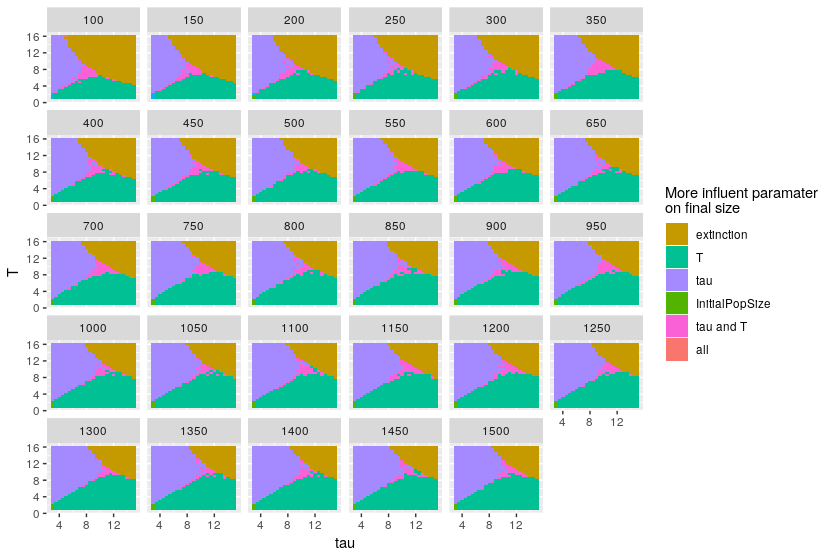} 
     \caption{Parameters having the greatest influence on final size. Each plot corresponds to a fixed value of $InitialPopSize$ specified above the plots (from 100 to 1500 crowns), $\tau$ varies on the $x$-axis, and $T$ varies on the $y$-axis.}
   \label{Figure_influence_size}
\end{figure}

\begin{figure}[h]
    \centering
     \includegraphics[ width=0.90\linewidth]{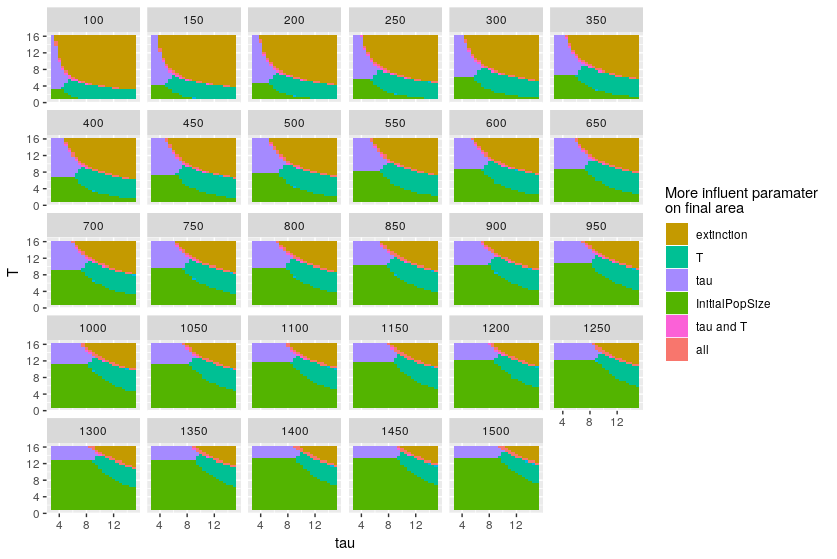} 
     \caption{Parameters having the greatest influence on final area. Each plot corresponds to a fixed value of $InitialPopSize$ specified above the plots (from 100 to 1500 crowns), $\tau$ varies on the $x$-axis, and $T$ varies on the $y$-axis.}
   \label{Figure_influence_area}
\end{figure}

Equations \eqref{eqRegressionGlobaleSize} and  \eqref{eqRegressionGlobaleArea} enable us to find which parameter most influences model outputs, and thus on which one it is better to concentrate management efforts.  To do so, we compared for each value of the triplet ($T$,$\tau$,$InitialPopSize$) the final size and area according to Equations \eqref{eqRegressionGlobaleSize} and \eqref{eqRegressionGlobaleArea},  for the three following parameter value combinations:  ($T+1$,$\tau$,$InitialPopSize$), ($T$,$\tau+1$,$InitialPopSize$) and ($T$,$\tau$,$InitialPopSize \times 0.9$).
Each plot on Figures \ref{Figure_influence_size} and \ref{Figure_influence_area} corresponds to a fixed value of  $InitialPopSize$, with $\tau$ varying on the $x$-axis and $T$ varying on the $y$-axis, and associates with each triplet the  most important parameter in a management perspective, that is the parameter the modification of which produces the lowest output (it happens that two modifications produce the lowest output).  Brown zones correspond to set of parameter values that lead to eradication, thus in this zone no gain can be expected by any modification. Figure \ref{Figure_influence_size} shows that, out of the extinction zones, $T$ or $\tau$ have the greatest influence on final size and allows to determine the most efficient management modification. Especially, when $\tau$ is low, it is more efficient in terms of size reduction to mow one more time each year, and conversely, when $T$ is low, it is more efficient to continue mowing one more year. As for the final area, we observe on Figure \ref{Figure_influence_area} that areas corresponding to a greatest influence of $T$ or $\tau$ are reduced compared to Figure \ref{Figure_influence_size}, in the favor of the area of greatest influence of $InitialPopSize$.
In these  regions of parameter values, beginning the management project on smaller stands (size equal to 90\% of reference size) has more impact on the final area than mowing one more time
each year or over a one year longer period of time;  thus in these conditions, early detection of stands should be  encouraged.

\section{Discussion}
\label{Section_Discussion}

In this paper, we proposed a stochastic individual based model for the growth of a stand of Japanese knotweed including mowing as a management technique. Then, we calibrated plant dynamics parameters with field data in Section \ref{Subsection_resultCalibration}. The set of parameters obtained was in agreement with values of parameters available in the literature and with our field experience.  In Sections \ref{Section_T} - \ref{Section_TailleIni} we stu\-died the influence of the initial population size, the mean number of mowing events a year and the management project duration on mean area and mean size of stands. We also obtained formulas for the area and size of a knotweed stand, as functions of those management parameters (for $\tau >2.5$) and initial population size. 
We have shown that mowing is not sufficient to decrease the population size and area. Indeed, if the number of mowing events per year is too small (smaller than 2.5 in our case), the population size and area increase during the management project. 
We have  also shown how those results could be used by managers.
Simulation results suggest the minimal duration of the management project necessary to achieve eradication (if it is possible at all, given a certain frequency of mowing).
In Figure \ref{Figure_QuadraticRegresionCurve}, we  plotted  the quadratic regression curves for the average final area  with respect to the duration of the management project ($T$), obtained for different values of $\tau$ and for given values of $InitialPopSize$ and $proportionMowing$. 
The figure indicates the potential benefit in terms of invaded area reduction, of mowing the stand once more each year. More generally, formulas  given in Equations \eqref{eqRegressionGlobaleSize} and \eqref{eqRegressionGlobaleArea}, highlighting the relation between final outputs, management parameters and initial population size, enable to answer questions about the efficiency of different mowing strategies.

Following \cite{dauer2013elucidating} and \cite{smith2007simulation}, we have assumed that the invasion occurs in an homogeneous area. Models that take into account the inhomogeneities of the invaded land are often static models, which do not consider the invasion as a dynamic phenomenon (no temporal component \citep{lookingbill2014incorporating, buchadas2017dynamic}). \cite{lookingbill2014incorporating} use indices such as habitat suitability, constructed from field data such as  humidity or soil type, to produce invasibility maps. These maps assign a score to each zone which describes its probability of being invaded.

Another simplification in our model is that we did not take into account the dispersal of fragments of rhizome due to mowing. This may however be an important way of propagation of the plant in some conditions and it contributes to its invasiveness \citep{sasik2006rhizome}. Dispersal has to be considered if one wants to model the invasion of Japanese knotweed at the scale of a region composed of several stands. This will be the subject of future work. We could formulate this problem in the formalism of the viability theory \citep{aubin1991viability}. 
In this framework, the dynamics of the system depends on the system state and on controls.
One objective is to prove the existence of controls and to find initial values of the system such that the system state remains in a set of constraints (e.g. the invaded area below a given threshold). For example, managers could be interested in controlling the density of Japanese knotweed. Then we could study the resilience of the system, that is to say its ability to recover a property after a perturbation.

The importance of integration (biomass transfer between crowns) is still under debate. In \cite{price2002seasonal}, the authors notice that there is relatively few integration, whereas in \cite{suzuki1994growth} the authors found a larger integration. We did not consider this process in our model.
Adding this phenomenon could produce simulation results closer to reality.

The field data we have used for the calibration were extracted from \cite{martin2018Biological}. The measurements were carried out in 2008 and 2015 on stands that were mown or not. For each stand, data provide some information about the management technique used by the land owner: the frequency of mowing and an estimation of the proportion of the mown stand. Calibration results for some of the plant dynamics parameters based on these measurements are in agreements with data found in the literature.

The model is written in the formalism of measure-valued stochastic processes. The tools we used here for the particular case of the management of a Japanese knotweed stand can be used in a more general context. In particular, we could apply this method to others invasive plants, like seed dispersal species. One could even allow individuals to move in such models. In \cite{leman2016convergence}, the author took into account the spatial motion in an individual-based stochastic population model.  Furthermore, including sexual reproduction of individuals, as in \cite{smadi2018looking}, would also enable to consider animal invasive species, like mosquitoes \citep{juliano2005ecology} or feral cats \citep{baker2016placing}.

\section*{Aknowledgements}

This work was partially funded by Electricité De France (EDF) and we thank Laure Santoni and Agnes Bariller for helpful discussions and comments. FL, SM and CS acknowledge partial funding from the Chair "Modélisation Mathématique et Biodiversité" of VEOLIA-Ecole Polytechnique-MNHN-F.X.
FL and BR also acknowledge partial funding through the ANR "Alien" project (ANR 14-CE36-0001-01).
Finally, IA, FL, SM and CS acknowledge Complex System Institute of Paris Île de France for the hosting, and the OpenMOLE team for their advice on the software.

\appendix

\setcounter{figure}{0} 
\renewcommand{\thefigure}{A.\arabic{figure}}

\setcounter{table}{0} 
\renewcommand{\thetable}{A.\arabic{table}}


\section{Summary of model parameters}  \label{Section_Summary_of_model_parameters}

Table  \ref{tableau_recap_param} focuses on the plant dynamic parameters. 
We precise parameters units for those that have a biological meaning.
\bigbreak 
\fontsize{6pt}{7pt}\selectfont
\begin{tabular}{lll}
	\hline
   Variable & Description    & unit                        \\
   	\hline
   Biomass &                              \\ 
   	\hline
    $K$       &  maximal biomass (Equation \eqref{solVonBert}) &  g  \\  
    $L$      &   biomass growth rate for low biomass  (Equation \eqref{solVonBert})  & $year^{-1}$  \\ 
    $a_0$      & initial biomass (of a crown at birth)   &  g \\ 
       	\hline
   Mowing &                         \\  	 \\
		\hline   
    $mowingParameter$  &  in the mowing effect function in Equation \eqref{fauche_lambda} & $g^{-1}$ \\  
          \hline
   Mortality &                         \\ 
		\hline
    $deathParameterScaling$  & mortality rate for the low biomasses in Equation \eqref{eqtaux_mortalite} & $year^{-1}$ \\ 
    $deathParameterDecrease$  & decay rate of mortality function in Equation \eqref{eqtaux_mortalite} & $g^{-1}$ \\
       \hline
  Birth &                           \\ 
		\hline
    $distanceParent$      &  apical dominance distance (Equation \eqref{eq_birth})  &  m \\ 
    $distanceCompetition$ &  intraspecific competition distance (Equation \eqref{ensemble_dispersion})  &  g  \\ 
 	$\bar{b}$    &  birth rate (under ideal conditions)     & $year^{-1}$  \\
    $ (\text{shape},\text{scale})$     &   Gamma law, dispersion of the new individual\\ 
     \hline
\end{tabular}
 \captionof{table}{Summary of model parameters}
 \label{tableau_recap_param}

\normalsize
\bigbreak 

We have the following management parameters:
\begin{itemize}
\item mean number of mowing events a year: $\tau$
\item management project duration: $T$ 
\item proportion of mown crowns: $proportionMowing$
\end{itemize}

and initial population size parameter: $ InitialPopSize $.


\section{Description of the algorithm used to simulate a solution of Equation  \eqref{equation_sto}. \label{Subsection_Description_algo}}

We present one step of the algorithm that enables to make evolve a stand under mowing.

At each time, there are three possible events: a birth, a death or the mowing of a proportion $proportionMowing$ of individuals in the population.
Suppose we have $N$  individuals at a time $t$.
We start by calculating the next time at which there is an event, which requires the sum of the rates of the events of birth, death and mowing). The law of this time only depends on the current population state as the process is Markovian.

If it is a birth event, we select the parent uniformly at random in the population. We check whether the individual selected to be the parent does not already have too many neighbours at a distance lower than $distanceParent$. If it occurs, we draw the position of the new individual (child) according to the Gamma law described in the paragraph "Dispersal of the created individual" of Section \ref{paragraph_DispersalOfTheCreatedIndividual} (with an angle chosen uniformly around the  potential parent). If the child's position falls into the set $C$ of Equation \eqref{ensemble_dispersion} (i.e. it does not fall into an intraspecific competition zone), then the individual is born at this position, and the new population size is $N +1$. Otherwise, i.e. if the parent has already enough crowns close to it, or if the new individual to be born is out of the set $C$, then there is no birth, and the population size remains $N$.

If the event is a mowing event, then we mow every indivi\-dual with probability $proportionMowing$: we replace its biomass $a$ by $ a*F(a)$.
 
Finally, if the event is a death event, then the individual likely to die is drawn uniformly at random, and we  choose  with the realization of a random variable whether this individual really dies according to its mortality rate, which depends on its biomass. If it dies, it is taken out of the population and the new population size is $N -1$, otherwise, nothing happens.

Between each event, the biomass of each individual grows in a deterministic way. The algorithm stops as soon as there is no more individual in the population.

\bigbreak 
For the simulations, we have to specify an initial population (at time 0, positions and biomass) that will evolve. To create an initial population, we use the algorithm above with initially one individual until the population reaches a prescribed size, with a low management technique ($\tau = 1$ and $proportionMowing =  0.9$). Indeed, if we put a larger value of $\tau$, we cannot often create the initial population because the original individual dies before producing offspring. In this spirit, we also let several attempts in the algorithm to create the initial population.


\section{Plots of functions defined in Section \ref{Section_Description_Phenomena} with plant dynamics parameters from the calibration}

\begin{figure}[!h]
    \centering
     \includegraphics[ width=0.90\linewidth]{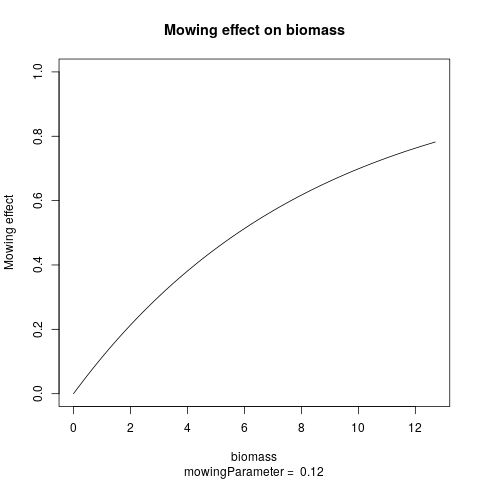}  
     \caption{Function for the effect of mowing, Equation \eqref{fauche_lambda} with parameter values from the calibration (Table \ref{tableau_result_Calibration}). }
   \label{Figure_MowingEffect}
\end{figure}

\begin{figure}[!h]
    \centering
     \includegraphics[ width=0.95\linewidth]{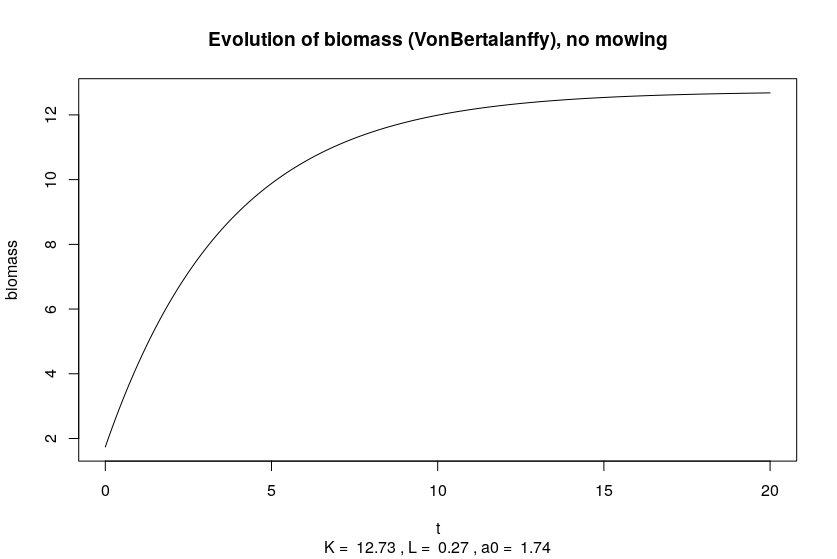}  
     \caption{Evolution of biomass, without mowing event, during 20 years (Equation \eqref{solVonBert}) with parameter values from the calibration (Table \ref{tableau_result_Calibration}).}     
   \label{Figure_Evolution_Biomass}
\end{figure}

\begin{figure}[!h]
    \centering
     \includegraphics[ width=0.90\linewidth]{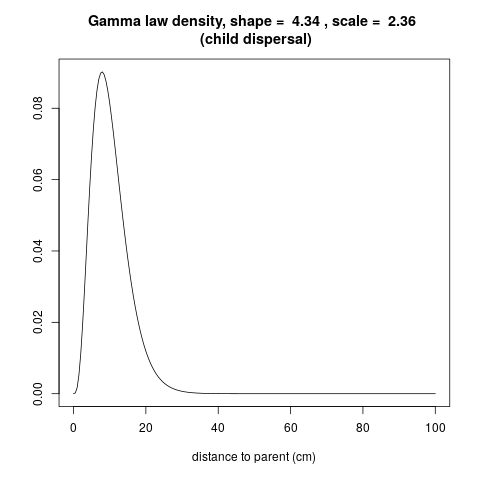}  
     \caption{Gamma law for the distance of dispersion with parameters values from the calibration (Table \ref{tableau_result_Calibration}). }     
   \label{Figure_GammaLaw}
\end{figure}

\begin{figure}[!h]
    \centering
     \includegraphics[ width=0.90\linewidth]{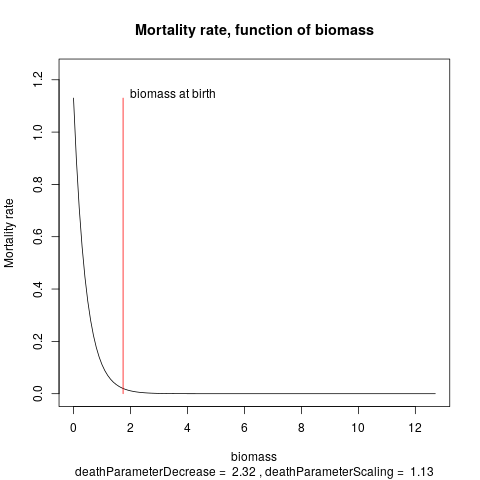}  
     \caption{Mortality rate, function of biomass (Equation \eqref{eqtaux_mortalite}) with parameters values from the calibration (Table \ref{tableau_result_Calibration}).}     
   \label{Figure_Mortality}
\end{figure}

\newpage

   \bibliography{biblio_article1_JTB_v3}

\end{document}